\documentclass[11pt]{article}
\usepackage{graphicx}
\usepackage{amsmath}
\usepackage{amssymb}
\usepackage{cite}
\usepackage{color} 

\usepackage{url}

\usepackage[labelfont=bf,labelsep=period,justification=raggedright]{caption}

\makeatletter
\renewcommand{\@biblabel}[1]{\quad#1.}
\makeatother

\date{}

\begin{document}

\begin{flushleft}
{\Large
\textbf{The role of asymmetric interactions on the effect of habitat destruction in mutualistic networks}
}
\\
Guillermo Abramson$^{1,2,\ast}$, 
Claudia A. Trejo Soto$^{2}$, 
Leonardo O\~{n}a$^{3}$
\\
\textbf{1} Centro At\'{o}mico Bariloche and CONICET, R8400AGP Bariloche, Argentina
\\
\textbf{2} Instituto Balseiro, R8400AGP Bariloche, Argentina
\\
\textbf{3} Department of evolutionary genetics, Max Planck Institute for Evolutionary Anthropology, Deutscher Platz 6, 04103 Leipzig, Germany
\\
$\ast$ E-mail: abramson@cab.cnea.gov.ar
\end{flushleft}

\section*{Abstract}
Plant-pollinator mutualistic networks are asymmetric in their interactions: specialist plants are pollinated by generalist animals, while generalist plants are pollinated by a broad involving specialists and generalists. It has been suggested that this asymmetric ---or disassortative--- assemblage could play an important role in determining the equal susceptibility of specialist and generalist plants under habitat destruction. At the core of the argument lies the observation that specialist plants, otherwise candidates to extinction, could cope with the disruption thanks to their interaction with generalist pollinators. We present a theoretical framework that supports this thesis. We analyze a dynamical model of a system of mutualistic plants and pollinators, subject to the destruction of their habitat. We analyze and compare two families of interaction topologies, ranging from highly assortative to highly disassortative ones, as well as real pollination networks. We found that several features observed in natural systems are predicted by the mathematical model. First, there is a tendency to increase the asymmetry of the network as a result of the extinctions. Second, an entropy measure of the differential susceptibility to extinction of specialist and generalist species show that they tend to balance when the network is disassortative. Finally, the disappearance of links in the network, as a result of extinctions, shows that specialist plants preserve more connections than the corresponding plants in an assortative system, enabling them to resist the disruption.

\section*{Author summary}
In plant-pollinator mutualistic networks, plant species are considered generalists when pollinated by several or many animal species of different taxa, and specialists if pollinated by one or a few taxonomically related pollinators. Theory predicts that specialized plants are more vulnerable to disruption induced by habitat destruction, because they cannot counterbalance for the loss of their few specific mutualistic partners with other alternative pollinators. Contrary to these theoretical expectations, specialist and generalist plant species have shown to be equally affected by habitat fragmentation. One explanation for such pattern relies on the asymmetry (disassortativity) of these networks, where specialist plants are mainly pollinated by generalist pollinators whereas generalist plants are pollinated by both specialist and generalist pollinators. Under habitat fragmentation, specialist plants would therefore be able to keep their few pollinators and thus their reproduction would not be so severely impaired as previously thought. In this work, we mathematically test this hypothesis and conclude that disassortative networks indeed favour a balanced extinction of generalist and specialist plants when subject to habitat destruction. The pattern is stronger under more sparse network connectivity, and higher levels of destruction.

\section*{Introduction}

Habitat destruction is the major cause of species extinctions and a main driving force behind current biodiversity loss \cite{Ehrlich:1981p1, Wilson:1988p1, Perrings:1995p1}. Animal-mediated pollination is crucial for the sexual reproduction of flowering plants, and one of the most actively studied process in the context of habitat fragmentation. The strength of the effect of fragmentation on pollination and on plant reproductive success shows a highly significant correlation, suggesting that one of the most important causes of reproductive impairment in fragmented habitats may be pollination limitation \cite{Aguilar:2006p968}.

In the mutualistic interaction between plants and pollinators, plant species are typically considered generalists when pollinated by several or many animal species of different taxa, and specialists if pollinated by one or a few taxonomically related pollinators  \cite{Bond:1994p83, Herrera:1996p65, Wasser:1996p1043, Renner:1999p339}. Most plant-pollinator mutualistic networks have shown to be highly asymmetrical with specialist plants being pollinated mostly by generalist pollinators, whereas generalists are pollinated by both specialists and generalists pollinators \cite{Bascompte:2003p9383, vazquez2004} .

Some ecological consequences of the asymmetry of the plant-pollinator mutualistic network have been studied. Using mathematical models it has been shown that the asymmetry of plant-pollinator networks differs from random networks in their response to habitat destruction. Networks with topologies present in real communities start to decay sooner than random communities, but persist for higher destruction levels. When the destruction level is above a given threshold the whole community collapses  \cite{Fortuna:2006p281}. 

Theoretical studies have suggested that habitat destruction would affect preferentially specialised plants, because they would not be able to counterbalance the loss of their few specific mutualist partners with other alternative pollinators \cite{Bond:1994p83, Wasser:1996p1043, Fenster:2001p844}. Generalist plants, instead, should be more adaptable to the changes imposed by fragmentation on their pollinator assemblages because the absence of one or some of their pollinators could be compensated by other pollinators from their wide assemblages \cite{Morris:2003p260}. Contrary to these theoretical expectations, no significant difference was found in the mean effect sizes of specialist and generalist plant species, both being equally negatively affected by habitat fragmentation \cite{Aizen:2002p885, vazquez2002}. One explanation for the equal susceptibility of specialist and generalist plant to habitat destruction is that because specialist plants interact mainly with generalist pollinators, they would be able to keep their few pollinators in fragmented habitats, and thus their reproduction would not be so severely impaired as previously thought. Generalist plants, which interact with both generalist and specialist pollinators, would tend to loose their specialist pollinator fraction from their assemblages and retain their generalist pollinators. Thus, a decrease in the remaining generalist pollinators population would therefore have equal effects on the two groups of plants \cite{Ashworth:2004p717}.

Mathematical models differ from verbal theories in giving a precise connection between assumption and conclusion. They are the key tool needed to illuminate how the network architecture influences species extinction or persistence \cite{Bascompte:2010p765}. In this work we constructed assortative and disassortative networks and analyzed the effect of habitat destruction in each case, focusing on the relative effect on specialist and generalist species. We find that the way in which species are interconnected determines in a great deal who gets extinct, and in which way the perturbation affects the balance of specialization. In accordance with the theory proposed by Ashworth \cite{Ashworth:2004p717}, we observe that in asymmetric (disassortative) networks, generalist plants loose their connections with specialist pollinators, but specialist plants loose by far much less connections than specialist ones in the symmetric (assortative) networks. Our results support the idea that network asymmetry explains the equal susceptibility of generalist and specialist plants to habitat disturbance.

\section*{Models}

\subsection*{Interaction networks}

\begin{figure}[th]
\begin{center}
\includegraphics[width=3in]{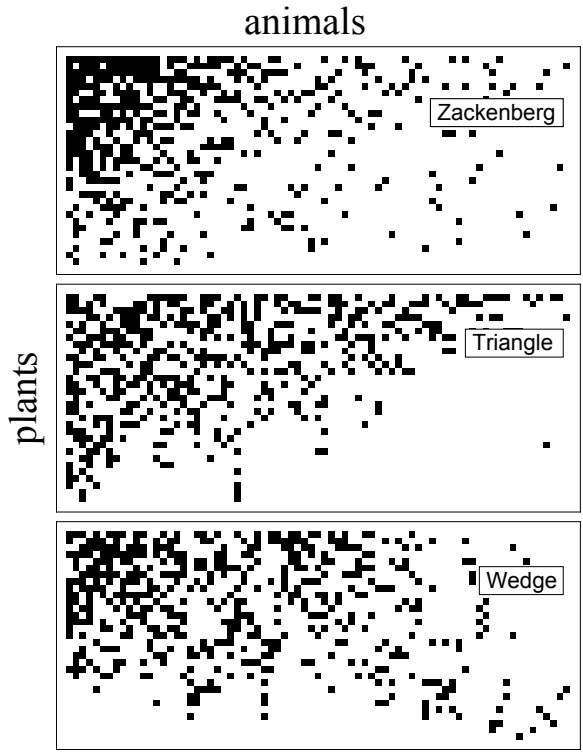}
\end{center}
\caption{{\bf Interaction networks of mutualistic systems.} 31 plants (rows), 75 animals (columns). A black dot represents an interaction between the corresponding plant and animal. Top: Zackenberg network, $r=-0.1$, $\eta=0.38$. Center: triangle model, $r=-0.38$, $\eta=0.38$. Bottom: wedge model, $r=0.30$, $\eta=0.24$. Density of links is $\rho=0.195$ in all three systems.} 
\label{redes}
\end{figure}

To test the theoretical validity of this hypothesis we propose to analyse the dynamics of mutualistic systems based on several different models of interaction. We can manipulate these models in ways that cannot be done in natural systems, a fact that provides a good testbed for the construction of theoretical hypotheses and predictions. We will analyze two families of interaction networks, based on prototypical topologies, that range from those similar to natural ones (asymmetric) to their opposite (symmetric). We call these families the \emph{triangle} and the \emph{wedge} models. Imagine the interaction matrix of a system with $N$ plants and $M$ animals as a rectangular matrix with the plants arranged in rows and the animals in columns, all of them sorted from generalist to specialists. Matrix elements $q_{ij}$ indicate the existence (1) or absence (0) of interaction between plant $i$ and animal $j$. A triangle has interactions according to:
\begin{align}
q_{i,j}=\begin{cases}
         1& \text{if } j \le M-(M/N)\,i,\\
		 0& \text{otherwise},
		 \end{cases}
\end{align}
that is, a triangle of 1's occupying the upper-left half of the matrix. Such a system has a density of links $\rho\approx 0.5$. For the wedge family of models, let us define the connections as:
\begin{align}
q_{i,j}=\begin{cases}
         1& \text{if } j \le M/2-M/(2N)\,i \\ 
		  & \text{and } j\ge -M+(2M/N)\,i,\\
		 0& \text{otherwise},
		 \end{cases}
\end{align}
that is, a ``kite'' figure with its narrow angle pointing towards the specialists-specialists zone. Both models, as defined, have densities of links much higher than what is usually observed in nature. Yet, they can be easily modified to achieve a prescribed density by turning a fraction of the 1's into 0's. Figure \ref{redes} shows a real network, Zackenberg station\protect\footnote{This is an arctic tundra ecosystem in Greenland, credited to H Elberling and J M Olesen and reported in \cite{Bascompte:2006sup}.}, together with a triangle and a wedge of the same density. 

The relevant topological characteristic of these networks can be quantified by three parameters. The density of links $\rho$, already mentioned, defined as the number of present links divided by the total number of possible links between plants and animals ($N\times M$). The asymmetry of the connections typical of mutualism, namely the fact that specialists interact with generalists and viceversa, is readily captured by the \emph{degree assortativity} (the degree of a node is the number of interactions it has). In the theory of complex networks, a network is called \emph{assortative} if the connections are mainly between nodes of similar degrees, and \emph{disassortative} if the connections are between nodes of very different degree (see for example \cite{Newman2010}, chapter 7). Natural mutualistic networks are disassortative. This is referred to as \emph{asymmetry} in the literature about mutualistic networks, and we use both terms as synonyms. The assortativity is measured by a coefficient $r\in(-1,1)$ \cite{Newman2002,Newman2010}. Networks with $-1<r<0$ are disassortative (asymmetric) and those with $0<r<1$ are assortative (symmetric).

Mutualistic networks are also highly \emph{nested} \cite{Bascompte:2003p9383}: the partners of a species of a given specialization level tend to be a subset of the partners of species with lower specialization. There isn't in the literature a consensus about the best way to measure nestedness in a quantitative way. We have chosen to use the one proposed by Almeida-Neto et al. \cite{Almeida-Neto2008}, which has good statistical properties, captures all the features of the concept of nestedness and requires no external parameters. The nestedness $\eta$ ranges from 0 to 1 for increasingly nested networks.

\begin{figure}[th]
\begin{center}
\includegraphics[width=4in]{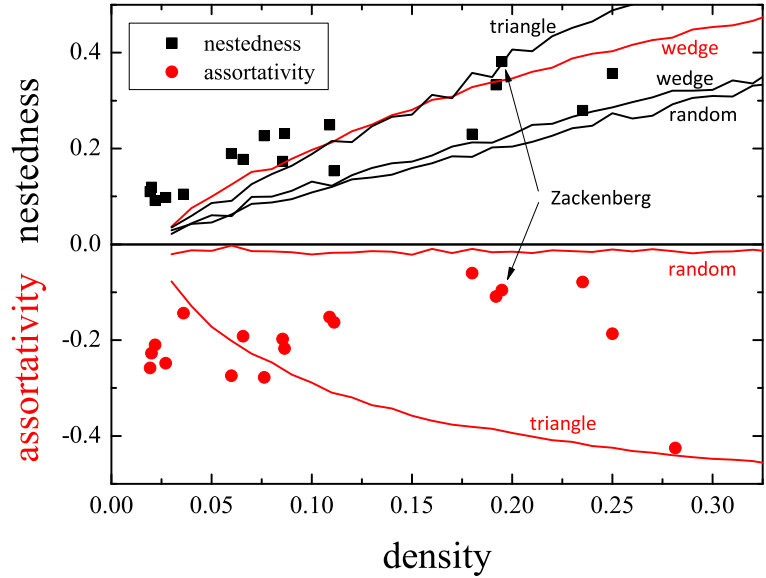}
\end{center}
\caption{{\bf Assortativity (red) and nestedness (black) plotted as a function of density of links.} The lines show the behavior of three models (triangle, wedge and random), while the points correspond to the real networks reported in \cite{Bascompte:2006sup}.} 
\label{parameters}
\end{figure}

Networks built with the triangle model are highly nested and very disassortative ($r<0$). Networks of the wedge family are moderately nested and very assortative ($r>0$). Figure \ref{parameters} shows a plot of the assortativity and the nestedness of both triangles and wedges as a function of the density of links. A null model consisting of random links is also shown, as well as a set of real networks (from  \cite{Bascompte:2006p431}). We can see that the triangle and the wedge models are clearly separated in assortativity: the triangles are disassortative and the wedges are assortative. They approach each other at low densities. Observe that several natural networks have very low density and very negative assortativity, a feature that is not captured well by the present models\footnote{We also implemented a Montecarlo algorithm to produce networks of arbitrary assortativity; the results will be presented elsewhere.}. We also observe that natural networks are more nested than the triangles. Both features, certainly, are due to the fact that natural networks are not built at random, but arise instead as the result of dynamics and evolution. We will address some of these questions below. 

\subsection*{Dynamical model of the mutualistic system}

We study the population dynamics of the mutualistic system by means of a model based on the Levins model for metapopulations \cite{Levins:1971p1246,tilman} . The destruction of the habitat is modelled as in \cite{tilman94}, with a single parameter $d$. Let us say that there are $N$ plants and $M$ animals in the system, and let us call $p_i$ and $a_j$ the population densities of plants and animals respectively. The evolution of these densities obeys the following dynamic equation (similarly to those proposed in \cite{Fortuna:2006p281}):
\begin{align}
\dot p_i &= p_i \sum_{j=1}^M q_{ij} r^a_{j} a_j (1-d-p_i) - m^p_i p_i, &i=1\dots N, \label{eq:plants}\\
\dot a_j &= a_j \sum_{i=1}^N q_{ij} r^p_{i} p_i (1-a_j) - m^a_j a_j, &j=1\dots M, \label{eq:animals}
\end{align}
where $r^p$ and $r^a$ are interaction intensities for plants and animals respectively, while $m^p$ and $m^a$ are death rates also for plants and animals. There are several simplifications in the interaction terms, that we have preferred to more involved ones to keep the number of parameters reasonably low and allowing us to concentrate on the effect of the network topology. Also, in order to keep the analysis centered on these matters, we have used values of the parameters drawn at random from uniform distributions. 
In this regard let us mention, nevertheless, that the intensities of interaction play an important role in the resistance to extinction. We have observed that if very small values of $r_i^{a(p)}$ are allowed in the system, the fraction of extinct species is considerably large. We have used an arbitrary value of $0.001$ as a lower cutoff for the values of $r^{a(p)}$, that permits the survival of a sizable system amenable for analysis. On the other hand, the mortality rates do not play a significant role in the extinction dynamics. The parameter that accounts for the destruction of the habitat is $d$, as mentioned. Observe that it affects only the dynamics of the plants, reducing their carrying capacity in (\ref{eq:plants}). With this, we are supposing that the destruction affects the available space for the sessile members of the community (through actual destruction, fragmentation, etc.), while the mobile pollinators are not directly affected by it. Of course, they feel the destruction indirectly by its effect on the plants populations. 

A realization of the dynamics starts from an initial condition consisting in random populations of plants and animals, and with $d=0$. As it is easy to imagine, a system with a random assemblage of plants and animals and random interactions between them behaves very differently from a natural one. A transient, during which some species get extinct, time needs to be discarded before. When the dynamics reaches a stationary state we take it as the initial state of the analysis. The network is smaller, as a result of the random extinctions, but we check that its size and its topological properties stay within a 10\% of the initially defined. For this system, a sudden destruction of a fraction of the habitat is simulated by setting a value of $d>0$.\footnote{We have also analyzed a scenario with progressive destruction, adiabatically increasing from 0 in small steps, and allowing for equilibrium to be reached before proceeding to further perturbation. This will be reported elsewhere.} As a result of the perturbation, additional species get extinct until a new stationary situation is achieved. At this moment our simulation ends, and we proceed to measure the properties of the system. The results reported below correspond to statistical averages over initial conditions, network connectivities and intensities, as indicated in each case. The only different situations correspond to the analysis of real networks, where the averages run only on initial conditions and parameters (but not on the network itself). 

\section*{Results}

\subsection*{Extinctions in the perturbed system}

\begin{figure}
\begin{center}
\includegraphics[width=4in]{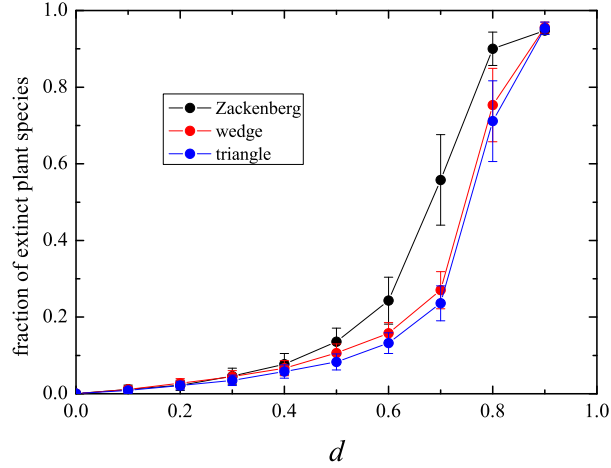}
\end{center}
\caption{{\bf Fraction of extinct plants as a function of the destruction parameter, in two network models and a natural network.} Links density is $\rho=0.2$ in the models and $\rho=0.18$ in the Zackenberg one. Average of 1000 realizations.} 
\label{fig:extinction}
\end{figure}

\begin{figure}
\begin{center}
\includegraphics[width=4in]{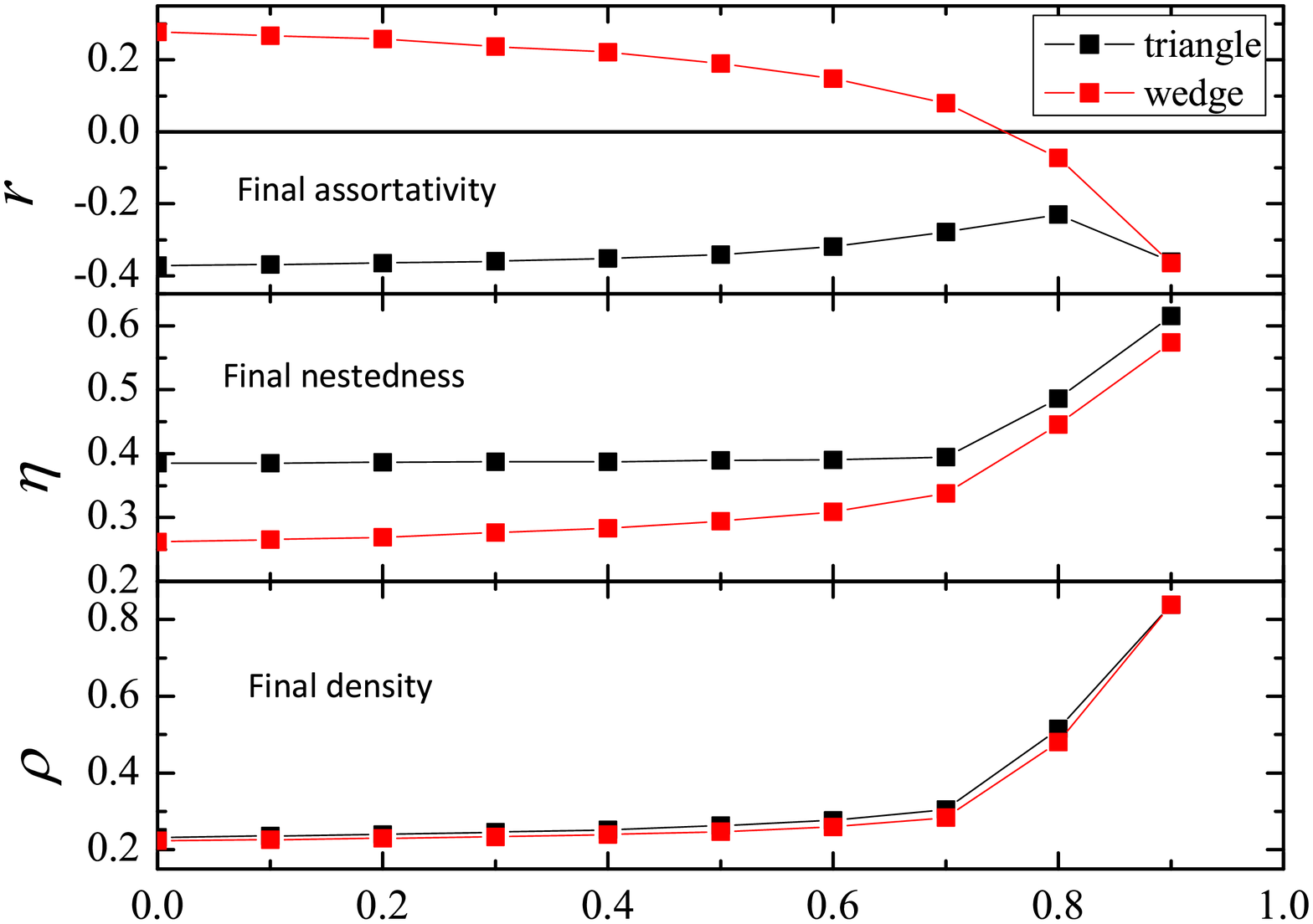}
\end{center}
\caption{{\bf Final network properties as a function of the destruction parameter.} Average of 1000 realizations for model networks with $\rho=0.2$ initially.} 
\label{fig:network}
\end{figure}

Let us briefly explore the general behavior after the perturbation, before proceeding to the matter of extinctions by specialization degree. Figure \ref{fig:extinction} shows the fraction of extinct species of plants as a function of the destruction parameter $d$. It is seen that a picture similar to the critical behavior explored in \cite{Bascompte:1996} arises: the extinction climbs steeply when a critical value of $d$ is approached. Since the size of the analyzed systems is finite the behavior is smooth, rather than abrupt as in a critical transition. Figure \ref{fig:extinction} shows a comparison between three networks of similar density: Zackenberg station, a triangle and a wedge. Observe how the real network is slightly more vulnerable than the artificial ones. When a system settles in its new equilibrium, not only a fraction of plant and animal species have got extinct, but also the network of interaction and its topological properties have changed. Figure \ref{fig:network} quantifies the effect on the three relevant parameters already introduced: assortativity, nestedness and density of links (for triangle and wedge networks of initial density $\rho=0.2$). Until $d$ gets very high (above 0.7, which implies a drastic modification of the system, as seen in Fig. \ref{fig:extinction}), the density of links changes very little in both the disassortative and the assortative systems. On the other hand, it can be seen that the main modifications are suffered by the wedge network, which evolves towards a state of higher nestedness (a little faster than the triangle) and, more importantly, of lower assortativity. In other words, the dynamics of extinction drives the assortative network (the symmetric system) towards a disassortative state, a state of asymmetric interaction. The relevance of this fact on the \emph{origin} of the observed asymmetric assemblage of natural mutualistic systems is without doubt an interesting one to be explored further, within a framework of evolving systems that lies beyond the present analysis.

\subsection*{Differential extinction by specialization degree}

We have seen that systems with very different topologies react pretty much in a similar way, regarding their loss of diversity, to the perturbation modelled as a destruction of habitat. In this section we show that, despite this global similarity, the way in which species are interconnected determines in a great deal \emph{who} gets extinct, and how the perturbation affects the balance of specialization. We want to show the differential effect of the extinction on generalist and specialist species. For this purpose we divide the totality of species (either plants or animals) in thirds according to their degree. We have three groups, then, and we call those most connected the \emph{generalists}, those less connected the \emph{specialists}, and the ones in the middle just so. The chose of just three groups is arbitrary, and has been preferred to having two groups in order to separate more clearly generalists and specialists. More than three groups are of course possible but,  for our purposes, not much is gained in increasing the resolution in specialization degree.

\begin{figure}
\begin{center}
\includegraphics[width=4in]{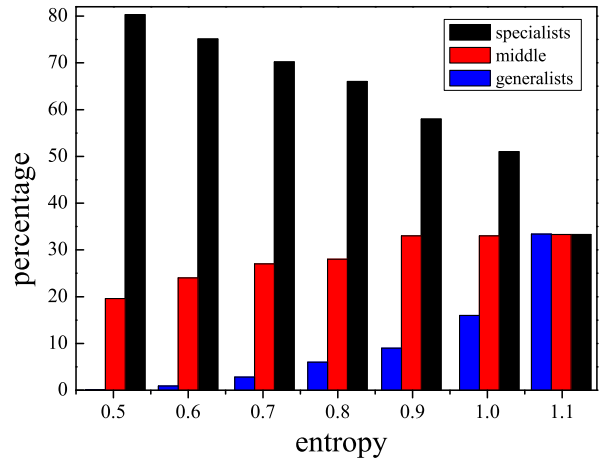}
\end{center}
\caption{{\bf Typical distributions and the entropies that measure their degree of uniformity.} }
\label{fig:distributions}
\end{figure}

Now, the question is how to measure the differential effect between generalists and specialists. Namely, how to distinguish between situations in which the extinct species belong equally to the three classes (no advantage in being a generalist) from situations in which the extinct belong substantially to the specialist class? A reasonable measure of this effect is provided by an entropy, defined as follows. Suppose that $n$ species get extinct at the end of the simulation, and that $n_1$ of them are specialists, $n_2$ are in the middle, and $n_3$ are generalists (so that $n=n_1+n_2+n_3$). Define:
\begin{align}
S=-\sum_{i=1}^3 p_i \log_{10} p_i,
\end{align}
where $p_i=n_i/n$. The entropy $S$ can range from 0 to $\log_{10} 3\approx 1.1$. This highest value corresponds to a uniform distribution of extinctions in the three specialization classes. Lower values correspond to non-uniform distributions. Figure \ref{fig:distributions} shows a few distributions and the corresponding values of entropy, as an example. 

\begin{figure}
\begin{center}
\includegraphics[width=4in]{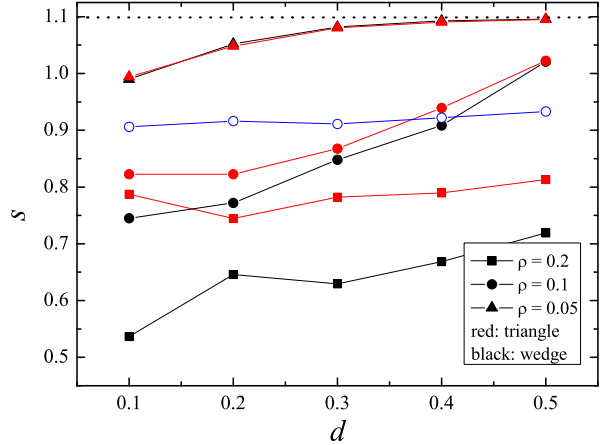}
\end{center}
\caption{{\bf Entropies of the distribution of extincts, as a function of habitat destruction.} Red: triangle model; black: wedge model. Three network densities are shown: low ($\blacktriangle$), middle ($\bullet$) and high ($\blacksquare$). The dashed line near the top represents the maximum entropy, corresponding to a uniform distribution. The blue line corresponds to simulations performed on the Zackenberg network (random parameters and initial conditions, without changing the network except for the transient).} 
\label{fig:entropies}
\end{figure}

Figure \ref{fig:entropies} shows the entropy as a function of habitat destruction for several system topologies. Each point in the plot corresponds to the average of 1000 realizations. First, observe that the entropy grows with $d$. This means that greater habitat disruptions produce more uniform extinctions. Observe also that (for each symbol type, representing different network densities) the triangle model (disassortative) displays higher entropy than the wedge one (assortative). The three densities used also show that this effect is more pronounced in high density system. This suggests a possible observation to be made in field studies: that in highly connected systems, asymmetric networks should display a balance (high entropy) in the resistance to extinction between generalists and specialists, while less asymmetric ones should  show a preferential extinction of specialists (lower entropy).

On the other hand, it is seen that systems with low density appear close together and close to the null model (corresponding to a random elimination of species, without dynamical evolution, and which gives perfectly uniform distributions, $S=\log 3$ for all $d$).

Finally, observe that the Zackenberg station system (density $\approx 0.2$) has a very high entropy, and a very flat dependence on $d$. This is another indication that the natural assemblage is not randomly assembled, but evolved instead. In this case, the consequence is that the natural network shows a significant balance between generalists and specialists. Bear in mind, however, that the dynamics mounted on the Zackenberg networks is simulated, and need not represent the dynamics of the real system; it is a dynamical property of just its network that comes into view in the present analysis.


To summarize, the entropy of the specialization degree of extinct species shows that, even if specialists are more susceptible to extinction, the distribution of degrees is flatter in the triangle model (asymmetric, disassortative, closer to natural) than in the wedge model (symmetric, assortative, far from natural). In other words: generalist and specialist are more equally susceptible to extinction in disassortative networks. The hypothesis of protection of specialists in disassortative networks (refer to Fig. 1 in \cite{Ashworth:2004p717}) claims that the degree of the partners plays an important role: in asymmetric interactions, specialists are protected because they are connected to generalists. The degree of the neighbors is \emph{precisely} what defines the assortativity of a network, so the behavior observed in our models supports that hypothesis. 

\begin{figure}[th]
\begin{center}
\includegraphics[width=4in]{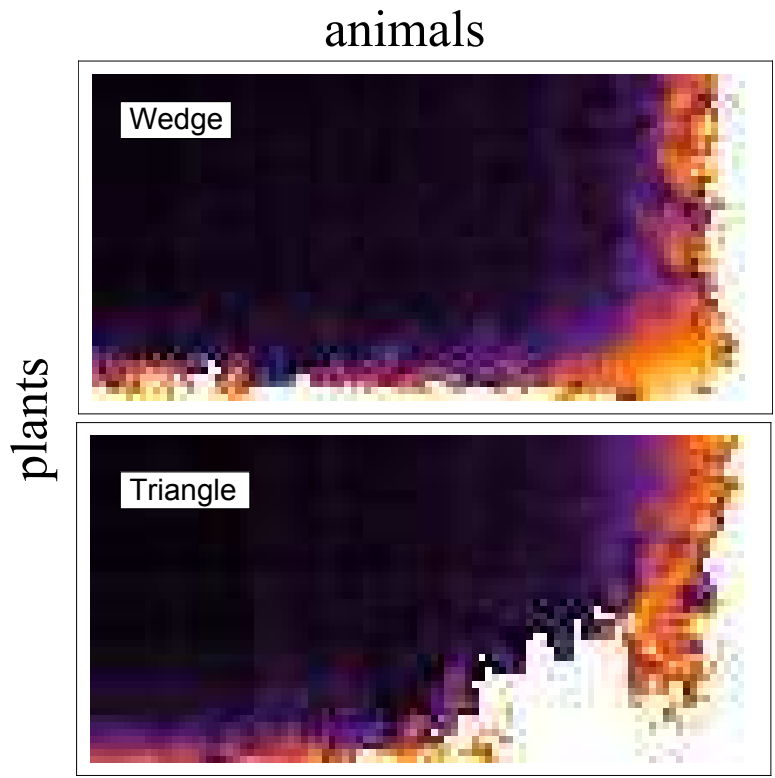}
\end{center}
\caption{{\bf Fraction of removed links.} These density plots represent the assemblage of plants (vertically) and their pollinators (horizontally), with generalists at the top and left, and specialist at the bottom and right, as in Fig. \ref{redes}. The colors code for the fraction of removed links when species get extinct as a result of the dynamic. Top: wedge model. Bottom: triangle model. The results have been averaged with a $3\times 3$ window to reduce fluctuations, and correspond to 1000 realizations each.} 
\label{linkext}
\end{figure}

We can further test this argument in the following way. Imagine not the extinct species but the \emph{links} that disappear in the system when a species goes extinct. What is the probability that a particular link disappears in the system? This can be computed by running many realizations of a particular model, and counting the fraction of links that disappear as a consequence of species extinction, at each place of the connection matrix sorted according to the degree of specialization. Figure \ref{linkext} shows this for the wedge and the triangular models. The colors of these density maps code for the probability of extinction of links, from 1\% to 50\%. 

First, observe that in both cases, the majority of extinct links occur along an envelope of the corresponding connection matrices. This means that, for species at each degree of specialization, it is more likely to loose their connections with specialist partners. This fits the expectations. 

Second, observe that (also in both models) more specialists get extinct (the probability increases towards the right and the bottom of the maps). This is also in agreement with expectations. 

Third and foremost, observe that the fraction of links that are lost in a wedge network is very high in the region of specialist-specialist region. On the contrary, the specialists in the triangle (connected mostly with generalists) have a darker color. They have a lower probability of extinction than the specialists connected to specialists in the assortative network. In other words: in the asymmetric system generalist plants loose their connections with specialists, but specialist plants, even when they loose connections, they loose much less ones than the corresponding specialists of the assortative network. This is precisely the phenomenon expected by the hypothesis of Ashwoth et al. \cite{Ashworth:2004p717}.

\section*{Discussion}

Mathematical models of plant-pollinator interaction networks have given many insights into the effect of habitat fragmentation on ecological communities \cite{Fortuna:2006p281, Bascompte:2010p765}. One of the main characteristics of the topology of plant-pollinator interaction networks is their asymmetry: specialist plants are mainly pollinated by generalist pollinators whereas generalist plants are pollinated by both specialist and generalist pollinators \cite{Bascompte:2003p9383, vazquez2004}. Such asymmetric type of interaction could be the reason why specialist and generalist plant species show similar response to habitat fragmentation \cite{Ashworth:2004p717}. The main aim of this work has been to test this hypothesis while giving it a theoretical framework. To this goal, we have constructed symmetric and asymmetric networks of plant-pollinator interactions (Fig. \ref{redes}). We have calculated the degree of asymmetry of such networks, as well as real ones, expressed by the measurements of nestedness and assortativity (Fig \ref{parameters}). We then analyzed the extinction pattern of these networks as a function of the degree of disturbance (Fig. \ref{fig:extinction}). We have also analyzed the nestedness, assortativity and density of the networks resulting from different degrees of habitat destruction (Fig. \ref{fig:network}). We have introduced entropy as a measure of the differential effect of habitat fragmentation on generalist and specialist species (Fig. \ref{fig:distributions}). Most importantly we have found that both the degree of connectivity and the degree of habitat fragmentation are factors that increase the pattern of equal susceptibility to generalist and specialist plant species to habitat destruction (Fig. \ref{fig:entropies}). A deeper analysis of the pattern of species extinction in symmetric and asymmetric networks shows that, in asymmetric (disassortative) networks, generalist plants loose their connections with specialist pollinators, but specialist plant loose by far much less connections than specialist plants in the symmetric (assortative) networks (Fig. \ref{linkext}). Therefore, and in accordance with Ashworth \cite{Ashworth:2004p717}, our results suggest that network asymmetry explains the equal susceptibility of generalist and specialist plants to habitat disturbance.

Our approach is similar than the one from Fortuna  \cite{Fortuna:2006p281} in that it does not include obligatory interactions on plants nor pollinators. We have assumed a community of facultative species in which the absence of their interacting partner does not implies species extinction. Obligatory interaction such as the one present in self-incompatible plants, may have a role on the pattern of species extinctions  \cite{Aguilar:2006p968}. We did not include other complex features in our model such as temporal variation in the association plant-pollinator \cite{Olsen:2008p1573} or spatial effects \cite{Fortuna:2008p490}. These processes can have a role in the response to habitat destruction and deserve further investigation. Our aim was to capture, with the simplest model, the effect of asymmetry on the pattern of extinction in response to habitat destruction.

\section*{Acknowledgments}
This work has been partially funded by CONICET through PIP 112-200801-00076 and by Universidad Nacional de Cuyo (06/C304). The funders had no role in study design, data collection and analysis, decision to publish, or preparation of the manuscript.

\bibliography{network}

\end{document}